\documentclass[11pt,a4paper]{article}
\usepackage{jheppub}
\usepackage[T1]{fontenc}
\usepackage{lmodern}
\usepackage{booktabs,mathtools,microtype,placeins}
\usepackage{flafter}
\renewcommand\afterProceedingsSpace{\vskip18pt}
\renewcommand\afterTitleSpace{\vskip18pt}
\renewcommand\afterRuleSpace{\vskip18pt}
\renewcommand\afterAuthorSpace{\vskip5pt}
\renewcommand\afterAffiliationSpace{\vskip3pt}
\renewcommand\afterEmailSpace{\vskip18pt}
\renewcommand\afterAbstractSpace{\vskip14pt}
\renewcommand\afterKeywordsSpace{\vskip14pt}
\newcommand{\dd}{\mathrm d}
\newcommand{\cG}{\mathcal G}
\newcommand{\cO}{\mathcal O}
\newcommand{\ent}{\mathcal S}
\newcommand{\energy}{\mathcal E}
\newcommand{\free}{\mathcal F}
\newcommand{\rsym}{r_{\rm sym}}
\newcommand{\etac}{\eta_{\rm c}}
\newcommand{\ww}{\mathfrak{w}}
\title{Entropically favored scalar hair on AdS black branes}
\author{Alex Buchel}
\affiliation{Department of Physics and Astronomy, University of Western Ontario,
London, Ontario N6A 5B7, Canada}
\emailAdd{abuchel@uwo.ca}
\abstract{We construct neutral AdS$_5$ black branes with source-free scalar hair
that is entropically favored over the symmetric phase. A quadratic
scalar--Gauss--Bonnet coupling destabilizes the thermal state while leaving
the ultraviolet operator dimension fixed at $\Delta=3$.
The fully backreacted solutions require only a quadratic scalar potential.
We describe how a neural-network search led to this simple model.}
\hypersetup{pdftitle={Entropically favored scalar hair on AdS black branes},pdfauthor={Alex Buchel}}
\begin{document}
\maketitle
\clearpage

\section{Introduction}
\label{sec:intro}
Conformal order is spontaneous symmetry breaking in a thermal state of a
conformal field theory, with no external source for the order parameter.
The absence of an intrinsic scale allows such order to persist at all
positive temperatures. The constructions of \cite{Chai2020} motivated an early
holographic search \cite{BuchelThermal}: the ordered states should be
described by neutral AdS black branes with scalar hair.
The possibility of ultraviolet symmetry breaking was first suggested in
the holographic model of \cite{BuchelPagnutti}, which
provided the starting point for the conformal-order construction of
\cite{BuchelThermal}.
Field-theory evidence for thermodynamically preferred order has since
grown, including large-$N$ conformal gauge theories
\cite{Chaudhuri2021}, models with persistent discrete symmetry breaking
\cite{Chai2022}, and the recent analysis of stable thermal minima in the
biconical model \cite{Smolkin2026}.

On the holographic side, a general perturbative prescription constructs
source-free hair by enhancing nonlinear scalar interactions; see
section~2 of \cite{BuchelFate} and \cite{BuchelQuest}.
The resulting ordered black branes, however, were entropically
subdominant. The ordered phases can also be perturbatively unstable,
despite having positive specific heat \cite{BuchelFate,BuchelCompact}.
The higher-derivative models studied in \cite{BuchelHD} did not rectify
the entropy disadvantage.

Our purpose here is to give a simple explicit construction of entropically
favored scalar hair. We use a quadratic coupling to the Gauss--Bonnet
(GB) density and a correlated quadratic potential. The vacuum mass stays
above the Breitenlohner--Freedman (BF) bound, while thermal curvature
induces an instability. This is the scalarization mechanism familiar from
scalar--GB gravity \cite{Silva2018}; related holographic scalarization
with planar AdS$_4$ horizons was studied in \cite{Guo2020}.
Here a fully backreacted AdS$_5$ example realizes preferred conformal order
without higher scalar interactions. The nonminimal curvature coupling
evades the hypotheses of minimally coupled no-hair arguments.

The rest of the paper is organized as follows.
Section~\ref{sec:model} introduces the model and its physical motivation.
Section~\ref{sec:instability} determines the perturbative instability
threshold of the symmetric phase.
Section~\ref{sec:ordered} constructs the fully nonlinear ordered black
branes and compares their thermodynamics with the symmetric phase.
Section~\ref{sec:causality} discusses causality and stability.
Section~\ref{sec:ML} explains how machine learning guided the search for
the model. Technical details of the linear analysis, numerical
solutions, and neural-network implementation are collected in
appendices~\ref{app:linear}--\ref{app:ML}.

\section{The model and its motivation}
\label{sec:model}
We work in five bulk dimensions with a real scalar $\phi$ and set the AdS
radius to one:
\begin{align}
 I&=\frac{1}{16\pi G_5}\int \dd^5x\,\sqrt{-g}\,
 \left[R+12-\frac12(\partial\phi)^2+F(\phi)\cG-V(\phi)\right],
 \label{eq:action}\\
 \cG&=R^2-4R_{ab}R^{ab}+R_{abcd}R^{abcd}.
 \nonumber
\end{align}
Here $G_5$ is Newton's constant, $R$ is the Ricci scalar, $F$ is the scalar
coupling to the GB density and $V$ is the scalar potential. The simplest
model that works has
\begin{equation}
 F(\phi)=\frac{\eta}{2}\phi^2,\qquad
 V(\phi)=\left(60\eta-\frac32\right)\phi^2,
 \label{eq:functions}
\end{equation}
where $\eta$ is a dimensionless coupling. The theory has a $\mathbb Z_2$
symmetry $\phi\mapsto-\phi$.
Pure AdS$_5$ has $\cG=120$, so the scalar mass is
\begin{equation}
 m_{\rm UV}^2=V''(0)-120F''(0)=-3=\Delta(\Delta-4),
 \qquad \Delta=3.
 \label{eq:mass}
\end{equation}
Here primes on $F,V$ denote derivatives with respect to $\phi$.
The dual scalar operator $\cO_3$ has dimension $\Delta=3$.
The mass lies above the AdS$_5$ BF bound
$m^2\ge-4$ \cite{BF}.

For the homogeneous and isotropic backgrounds we use ingoing
Eddington--Finkelstein (EF) coordinates,
\begin{equation}
 \dd s_5^2=\frac1{x^2}
 \left[-a(x)\dd t^2+s(x)^2\dd\vec y^{\,2}-2\dd t\,\dd x\right],
 \qquad \phi=p(x),\qquad 0\le x\le1.
 \label{eq:metric}
\end{equation}
The boundary is at $x=0$ and the horizon is at $x=1$.
Primes on radial functions denote derivatives with respect to $x$.
The symmetric solution is $a=1-x^4$, $s=1$, $p=0$.
Its metric fluctuations obey the Einstein equations.
The $\mathbb Z_2$-odd scalar fluctuation decouples, but its equation
retains the curvature coupling:
\begin{equation}
 [\Box-m_{\rm eff}^2(x)]\,\delta p=0,\qquad
 m_{\rm eff}^2(x)=-3-72\eta x^8.
 \label{eq:effective-mass}
\end{equation}
Here $\Box$ is the scalar wave operator in the symmetric geometry.
On this background the Weyl tensor $C_{abcd}$ satisfies
$\cG=120+C_{abcd}C^{abcd}=120+72x^8$.
The correlated potential cancels the $\eta$-dependent AdS contribution,
while the Weyl curvature supplies the negative thermal mass correction.
The scalar potential is bounded below for $\eta>1/40$, which includes
all the ordered solutions reported below.

The negative thermal mass motivates this choice of model: a sufficiently
deep radial potential can destabilize the black brane without changing the
vacuum dimension. The BF theorem itself concerns AdS; the global thermal
wave equation, rather than a pointwise mass comparison, must decide
stability, to which we turn now.

\section{Instability of the symmetric phase}
\label{sec:instability}
\begin{figure}[t]
 \centering
 \includegraphics[width=.78\textwidth]{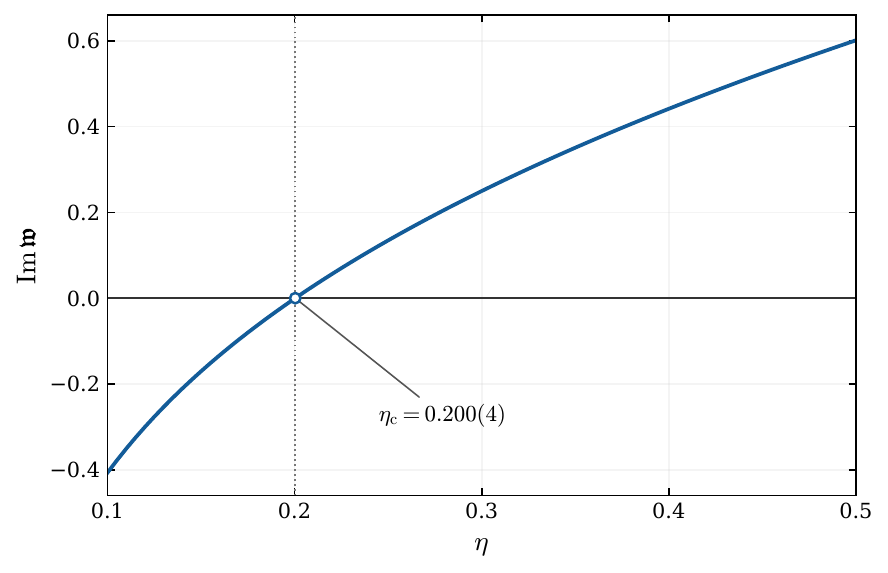}
 \caption{The homogeneous scalar quasinormal mode of the symmetric black
 brane, with $\ww=\omega/(2\pi T)$ and time dependence $e^{-i\omega\tau}$.
 The mode crosses zero at $\etac\simeq0.200(4)$.
 Negative and positive imaginary parts correspond to damping and growth,
 respectively. Each coupling $\eta$ specifies a different theory.}
 \label{fig:qnm}
\end{figure}

To study the scalar quasinormal spectrum, we introduce static, diagonal
Schwarzschild coordinates, related to the EF time
$t$ in \eqref{eq:metric} by
\begin{equation*}
 \tau=t+\int_0^x\frac{du}{f(u)},\qquad f(x)=1-x^4,
 \qquad
 \dd s_5^2=\frac1{x^2}
 \left[-f(x)\dd\tau^2+\frac{\dd x^2}{f(x)}+\dd\vec y^{\,2}\right].
\end{equation*}
The boundary times agree, while $\tau$ is singular at the horizon.
For a homogeneous mode $\delta p=e^{-i\omega\tau}\psi(x)$, we use the
physical dimensionless frequency
$\ww=\omega/(2\pi T)$, where $T$ is the Hawking temperature.
Since $T=1/\pi$ in these horizon units, the wave
equation is
\begin{equation}
 -\partial_x\left(\frac{1-x^4}{x^3}\psi'\right)
 -\frac{3+72\eta x^8}{x^5}\psi
 =4\ww^2\frac{\psi}{x^3(1-x^4)}.
 \label{eq:SL}
\end{equation}
The quasinormal boundary conditions are zero source, $\psi\sim x^3$,
and an ingoing horizon, $\psi\sim(1-x)^{-i\ww/2}$.
At onset, $\ww=0$, the horizon condition reduces to regularity,
$\psi'(1)=(3+72\eta)\psi(1)/4$.
The lowest, nodeless static mode occurs at
\begin{equation}
 \etac\simeq0.200(4).
 \label{eq:critical}
\end{equation}
Figure~\ref{fig:qnm} shows the scalar quasinormal mode connected to this
zero mode. It is purely imaginary over the displayed interval:
$\operatorname{Im}\ww<0$ below $\etac$, and
$\operatorname{Im}\ww>0$ above it. The mode therefore changes from a
damped fluctuation to an instability of the symmetric phase.
At fixed $\eta$, this damping or growth rate in units of $2\pi T$
is temperature independent.
Increasing $\eta$ lowers the radial quadratic form.
Finite spatial momentum adds a positive term, so the first onset is
homogeneous.

An analytic variational estimate also guarantees instability for
$\eta>1/3$ (appendix~\ref{app:linear}). The explicit nonlinear example
below uses $\eta=1/2$, safely inside this range.
The onset depends only on the quadratic coefficients of $F$ and $V$.
A local comparison of \eqref{eq:effective-mass} with the AdS BF value
is only diagnostic: the mass dips below $-4$ somewhere for
$\eta>1/72$, well before the global instability threshold
\eqref{eq:critical}.

\section{Fully nonlinear conformal order}
\label{sec:ordered}
We solve the fully backreacted equations from \eqref{eq:action},
with a regular horizon and source-free ultraviolet behavior
\begin{equation}
 a=1-\mu x^4+\cdots,\qquad s=1+\mathcal O(x^6),
 \qquad p=p_3x^3+\mathcal O(x^7).
 \label{eq:UV}
\end{equation}
Here $\mu$ sets the energy density and $p_3$ determines
$\langle\cO_3\rangle$. The absent non-normalizable $x$ term in $p$ imposes
zero scalar source; the absence of a term linear in $x$ in $s$ fixes the
residual radial shift. For $p_3\ne0$, the profiles $\pm p(x)$ describe
degenerate $\mathbb Z_2$-broken states.
Although the scalar equation is linear for a prescribed metric,
the metric responds to the scalar and makes the coupled problem nonlinear.
No quartic or higher scalar terms are needed for the solutions below.

We denote temperature by $T$ and entropy, energy and free-energy densities
by $\ent$, $\energy$ and $\free$, respectively. We set $16\pi G_5=1$,
so that $\energy=3\mu$.
For a planar horizon, the intrinsic-curvature correction to the Wald
entropy vanishes \cite{JacobsonMyers}. Therefore
\begin{equation}
 T=-\frac{a'(1)}{4\pi},\qquad
 \ent=4\pi s(1)^3,\qquad
 r=\frac{\ent}{T^3},\qquad
 \kappa=\frac{r}{\rsym},\qquad \rsym=4\pi^4.
 \label{eq:thermo}
\end{equation}
Note that we introduce $\kappa$ to distinguish the hairy black-brane phase
from the symmetric AdS--Schwarzschild phase, for which $\kappa=1$.
Flat boundary geometry and zero scalar source give the conformal relations
\begin{equation}
 \frac{\energy}{T^4}=\frac34 r,\qquad
 \frac{\free}{T^4}=-\frac14 r.
 \label{eq:CFT}
\end{equation}
The first equality is the conformal Smarr relation.
Thus, for $\kappa>1$, the hairy phase has larger $\ent/T^3$ and lower
$\free/T^4$ than the symmetric phase, at any temperature.

\begin{figure}[t]
 \centering
 \includegraphics[width=\textwidth]{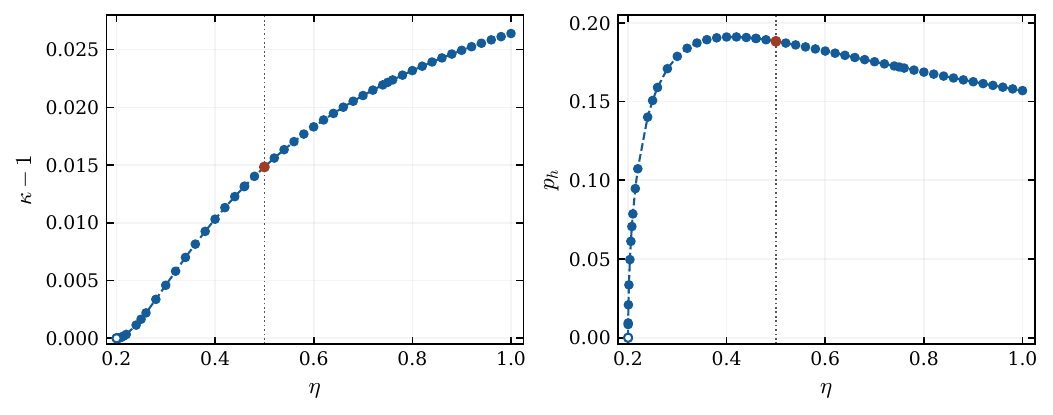}
 \caption{Ordered phases of AdS$_5$ black branes in the model
 \eqref{eq:action}--\eqref{eq:functions}.
 Left: the entropy-coefficient excess, $\kappa-1$.
 Right: the horizon scalar $p_h=p(1)$.
 The open point is the zero-amplitude onset \eqref{eq:critical};
 dashed segments guide the eye between samples.
 The red points and vertical dotted lines mark $\eta=1/2$.}
 \label{fig:branch}
\end{figure}

Figure~\ref{fig:branch} shows the sampled nodeless ordered branch.
All displayed samples have $\kappa>1$.
Varying $\eta$ changes the theory; it is not a temperature scan.
At any fixed coupling supporting an ordered solution, scale covariance
generates a thermal family with constant $\kappa$ and nonzero
$\langle\cO_3\rangle/T^3$.

\begin{figure}[!htbp]
 \centering
 \includegraphics[width=\textwidth]{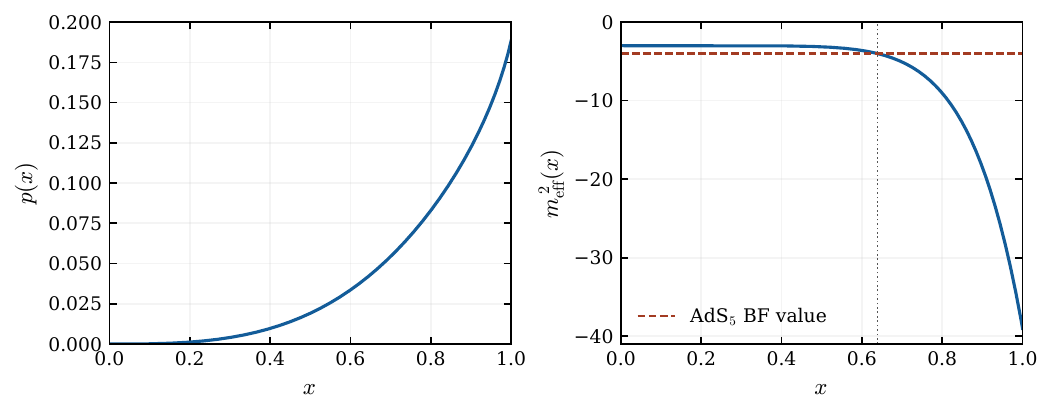}
 \caption{The model at $\eta=1/2$, highlighted in red in
 figure~\ref{fig:branch}.
 Left: the source-free scalar profile on the ordered black brane.
 Right: $m_{\rm eff}^2=-3-36x^8$ on the \emph{symmetric}
 AdS--Schwarzschild background.
 The horizontal dashed line is the AdS$_5$ BF value; it is crossed at
 $x=36^{-1/8}\simeq0.638(9)$.
 This local comparison illustrates the mechanism, while
 \eqref{eq:SL} determines the actual instability.}
 \label{fig:profile}
\end{figure}

A simple representative is $\eta=1/2$:
\begin{equation}
 F(\phi)=\frac14\phi^2,\qquad V(\phi)=\frac{57}{2}\phi^2.
 \label{eq:half}
\end{equation}
Its scalar profile is shown in figure~\ref{fig:profile}, together with the
symmetric-phase mass responsible for the instability.
The numerical solution gives
\begin{equation}
 \frac{r}{\rsym}=\kappa\simeq1.01(5),\qquad
 p_3\simeq0.149(7),\qquad p_h\equiv p(1)\simeq0.188(3).
 \label{eq:result}
\end{equation}
The entropy coefficient $\ent/T^3$ exceeds its symmetric-phase value by
$1.48(4)\%$ throughout the thermal family.
The entropy ratio follows from the numerical solution of the field
equations and is not imposed as a constraint.

\FloatBarrier
\section{Causality and stability}
\label{sec:causality}
In the symmetric phase, $F(0)=F'(0)=0$. The metric and scalar principal
parts are those of Einstein gravity and a canonical scalar, respectively.
Their characteristic cones are metric-null, even though the scalar mass
term produces the instability described above.

In the ordered phase, $F(p)\to0$ at the AdS boundary, removing the
constant ultraviolet GB deformation responsible for the familiar
near-boundary high-momentum causality constraint
\cite{BuchelMyers,BuchelCremonini}.
The scalar-dependent GB coefficient also remains small in a familiar
normalization. Using the conventional GB term
$\lambda_{\rm GB}\cG/2$ in the Lagrangian, define
$\lambda_{\rm GB}^{\rm eff}(x)=2F(p(x))=\eta p(x)^2$.
For \eqref{eq:half},
\begin{equation}
 \max_{0\le x\le1}\lambda_{\rm GB}^{\rm eff}(x)\simeq0.0177(2).
 \label{eq:lambda}
\end{equation}
This lies inside the constant-coupling causality interval
$-7/36\le\lambda_{\rm GB}\le9/100$
\cite{BuchelMyers,BuchelHoultKovtun}.
For a varying coupling this comparison is only indicative, since
derivatives of $F$ also enter the fluctuation equations
\cite{BuchelCremonini}.

Near $\etac$, the scalar hair can be made arbitrarily small, and the
ordered geometry approaches the causal AdS--Schwarzschild solution.
The scalar--GB coupling vanishes at the AdS boundary, and the tensor
propagation speed approaches the boundary speed of light from below.
These observations support our expectation that the theory remains
causal near onset.

We also examined helicity-two metric fluctuations of the ordered black
branes, following \cite{BuchelCremonini}. For all solutions shown in
figure~\ref{fig:branch}, the high-momentum effective potential, equal to
the squared propagation speed in boundary units, decreases
from one at the AdS boundary to zero at the horizon, excluding superluminal
propagation and eikonal instabilities. The tensor kinetic terms remain
positive, and the fluctuation equation admits no exponentially growing
modes at any real spatial momentum. The ordered phase is therefore causal
and linearly stable in this channel. The helicity-one and coupled
scalar--metric helicity-zero channels remain to be studied.

\section{How machine learning led to the model}
\label{sec:ML}
The simple model \eqref{eq:action}--\eqref{eq:functions} emerged from a
systematic search for entropically favored scalar hair. Using
physics-informed neural networks (PINNs), we explored the class of
theories \eqref{eq:action}, learning the scalar functions $F,V$ together
with the black-brane backgrounds.
Our implementation follows the PINN construction and numerical
verification strategy of
\cite{BuchelPINN}.\footnote{For earlier PINN reconstructions of holographic
theories from equations of state, see \cite{Bea2024,Jimenez2026}.}
Earlier neural approaches reconstructed bulk geometry from boundary data
\cite{Hashimoto2018AdS,Hashimoto2018QCD}, encoding discretized bulk
evolution in the network itself.
Here a prescribed entropy advantage guides a search for the bulk theory.

We parametrized the even functions as
\begin{align}
 F(\phi)&=\frac\eta2\phi^2+\phi^4\sum_{n=0}^{M}a_nT_n(2\phi^2-1),
 \nonumber\\
 V(\phi)&=\left(60\eta-\frac32\right)\phi^2
             +\phi^4\sum_{n=0}^{M}b_nT_n(2\phi^2-1),\qquad M=2.
 \label{eq:basis}
\end{align}
Here $T_n$ are Chebyshev polynomials of the first kind and $M$ is the
truncation degree. The basis is chosen for $|\phi|\le\Phi$, with $\Phi=1$.
The seven theory parameters $\eta,a_n,b_n$ were trained alongside the radial fields.
The ultraviolet mass condition was exact throughout.

The field parametrization imposed a zero scalar source, a regular horizon,
and $r=1.1\rsym$ as hard constraints. The loss minimized the regular
field-equation residuals, supplemented in the refinement by a Smarr
check and practical range and branch penalties
(appendix~\ref{app:ML}). The hard entropy condition excludes the symmetric
phase and ensures that any exact solution in this class has preferred
nonzero hair.

We then froze the learned $F,V$ and solved the boundary-value
problem with independent spectral fields, removing the entropy target.
The pilot and refined theories gave $\kappa\simeq1.05(0)$ and
$1.09(3)$, respectively, rather than exactly $1.1$.
The difference shows the importance of verifying approximate neural
candidates in the unconstrained problem.
Finally, at fixed $\eta\simeq0.459(2)$, we continuously scaled the
higher-order coefficients $a_n,b_n$ to zero while preserving $\kappa>1$.
Varying $\eta$ to $1/2$ then gave \eqref{eq:half}. The PINN search thus
identified theories supporting entropically favored scalar hair, and
numerical continuation reduced the learned theory to the simple quadratic
model. The final backgrounds and thermodynamic quantities were obtained
from numerical solutions of the field equations.

\acknowledgments
This work was supported by the Natural Sciences and Engineering Research
Council of Canada (NSERC).

\appendix
\section{The linear threshold}
\label{app:linear}
For the linear onset, $u=x^4$ and $\psi=x^3y(u)$ reduce the static
equation to
\begin{equation}
 u(1-u)y''+\left(\frac32-\frac52u\right)y'
 +\left(\frac92\eta u-\frac9{16}\right)y=0.
 \label{eq:linear-u}
\end{equation}
Primes in \eqref{eq:linear-u} denote $u$ derivatives.
Regular endpoint equations define a generalized eigenvalue problem.
Regular, source-free solutions determine $\etac\simeq0.200(4)$.
Returning to $x$, for nonzero frequency we solve the equivalent EF
equation, factoring
$\psi_{\rm EF}=x^3y(x)$:
\begin{equation}
 x(1-x^4)y''+(3-7x^4+4i\ww x)y'
 +(-9x^3+72\eta x^7+6i\ww)y=0.
 \label{eq:qnm-EF}
\end{equation}
Analyticity at both endpoints imposes zero source and an ingoing horizon.
For the purely imaginary branch, $\ww=i\gamma$, the equation is a real
generalized eigenvalue problem for $\gamma$.
Chebyshev collocation gives the curve in figure~\ref{fig:qnm}; the
crossing follows by varying $\eta$, without imposing a zero frequency.
Chebyshev degrees 48, 72 and 96 and endpoint-series matching are used to
control numerical errors.

There is also a useful analytic sufficient condition.
For the admissible trial function $\psi=x^3(1-x^4)^{1/4}$,
the numerator and denominator of the Rayleigh quotient are
\begin{equation}
 Q[\psi]=\frac{3\pi}{8}(1-3\eta),\qquad
 \|\psi\|^2=\int_0^1\frac{\psi^2\,\dd x}{x^3(1-x^4)}=\frac12.
 \label{eq:trial}
\end{equation}
Thus $\eta>1/3$ guarantees an unstable mode.
A negative eigenvalue gives $\ww=i\gamma$, $\gamma>0$.
The radial mode decays toward the horizon in Schwarzschild coordinates
and is regular in ingoing time.

\section{Numerical solutions}
\label{app:numerics}
We solve the scalar and metric equations from \eqref{eq:action}, including
the independent radial constraint. The equations are kept coupled in the
highest derivatives, and their
vanishing ultraviolet powers are removed analytically.
In particular, we do not divide by $a$ or by the radial principal
determinant, formed from the coefficient matrix multiplying the second
derivatives of the background fields.

We factor out the source and horizon behavior using smooth auxiliary
functions $N,S,Z$ and constants $P_0,H_0$:
\begin{equation}
 p=x^3[P_0+x^4N(x)],\quad
 s=e^{x^6S(x)},\quad
 a=(1-x^4)e^{x^4[H_0+x^2Z(x)]}.
 \label{eq:map}
\end{equation}
Thus $P_0=p_3$ and $H_0=1-\mu$.
We expand $N,S,Z$ in degree-$d$ Chebyshev
polynomials, with $P_0,H_0$ also unknown.
Four residual blocks are fitted at $2d+9$ Lobatto points, with analytic
polynomial derivatives. After removing ultraviolet powers, the metric-$s$,
metric-$a$, scalar and constraint residuals are divided by $6,24,12,24$,
respectively. We use the same normalization in appendix~\ref{app:ML}.
The $\eta=1/2$ convergence study below used
three-point coefficient Jacobians and tolerances $10^{-11}$.
The denser continuation branch in figure~\ref{fig:branch} used complex-step
coefficient Jacobians, tolerances $10^{-12}$ and degrees $56$--$104$,
with selected resolution checks through degree $120$.
The theory is fixed and neither the entropy ratio nor the Smarr relation
is imposed.

Two initial-guess routes reach the same $\eta=1/2$ solution:
continuation from the symmetric zero mode at nonzero scalar amplitude,
and removal of higher scalar interactions from a previously constructed
ordered solution. The latter initial guess was found in a
physics-informed neural-network search; the final quadratic-model
solutions require no neural representation.
The verification below concerns numerical solutions of the field equations.

\begin{table}[t]
 \centering
 \begin{tabular}{rccc}
 \toprule
 $d$ & $|\kappa_d-\kappa_{88}|$ & maximum scaled residual & Smarr residual\\
 \midrule
40 & $9.18(5)\times10^{-9}$ & $9.28(2)\times10^{-6}$ & $2.90(7)\times10^{-9}$\\
56 & $8.01(6)\times10^{-11}$ & $1.26(2)\times10^{-7}$ & $2.83(0)\times10^{-11}$\\
72 & $7.36(1)\times10^{-13}$ & $1.82(6)\times10^{-9}$ & $2.64(0)\times10^{-13}$\\
88 & $0$ & $2.62(5)\times10^{-11}$ & $2.88(7)\times10^{-15}$\\
 \bottomrule
 \end{tabular}
 \caption{Resolution convergence for the fixed theory \eqref{eq:half};
 $\kappa_d$ denotes the ratio $\kappa$ obtained at degree $d$.
 Residuals are evaluated on independent midpoint-angle nodes plus the
 endpoints. The Smarr residual is
 $\pi T s(1)^3-\mu$, where $\energy=3\mu$.
 Scaled equation residuals are numerical diagnostics, not error bounds.}
 \label{tab:convergence}
\end{table}

An independent dense-grid check confirms the residuals reported in
table~\ref{tab:convergence}. The scalar profile is monotone, and the
exterior geometry is regular.

\section{The constrained neural search}
\label{app:ML}
For the constrained search, three networks represent $N,S$ and an
auxiliary function $W$. We construct $Z$ in \eqref{eq:map} to impose the
target entropy ratio $\kappa_*$:
\begin{equation}
 Z(x)=S(1)-\frac13\log\kappa_*-H_0+(1-x)W(x),\qquad
 \kappa_*=1.1.
 \label{eq:hard}
\end{equation}
Indeed, writing $H=H_0+x^2Z$, one has $T=e^{H(1)}/\pi$,
$s(1)=e^{S(1)}$ and $H(1)=S(1)-\tfrac13\log\kappa_*$, so
$\kappa=e^{3[S(1)-H(1)]}=\kappa_*$ identically.
The entropy condition was built into the fields, rather than approximated
by a soft loss term. All networks used three hidden layers of width 32,
tanh activations and double precision. Further details of the PINN
implementation can be found in \cite{BuchelPINN}.

Denote the four scaled residuals defined in appendix~\ref{app:numerics},
in the same order, by $\widehat E_i$, $i=1,\ldots,4$.
The brackets $\langle\cdot\rangle$ denote
averaging over collocation points. The refined objective was
\begin{align}
 {\cal L}={}&\sum_{i=1}^4\langle\widehat E_i^2\rangle
 +\delta_{\rm Smarr}^2
 +10\langle[|p|-1]_+^2\rangle
 +10\langle[0.05-\widehat D]_+^2\rangle,\nonumber\\
 \delta_{\rm Smarr}={}&
 e^{4S(1)}\kappa_*^{-1/3}-(1-H_0),\qquad [z]_+=\max(z,0),
 \label{eq:loss}
\end{align}
where $\widehat D$ is the radial principal determinant normalized by its
AdS value. The last two terms enforce practical scalar-range and radial
branch preferences, not physical stability conditions.
The refinement used 192 jittered collocation points and a separate
1025-point audit. Geometry and theory learning rates began at
$10^{-4}$ and $3\times10^{-5}$ and decreased on plateaus.

The pilot used auxiliary derivative networks. Their consistency loss
could be small while the actual fields' second derivatives were
inaccurate; checkpoint selection therefore used actual-field residuals.
The refinement evaluated both first and second derivatives directly by
automatic differentiation. The best bulk loss,
$\sum_{i=1}^4\langle\widehat E_i^2\rangle$, was $1.3(1)\times10^{-5}$. Numerical solutions with the learned theory fixed removed
\eqref{eq:hard}, treating $Z$ independently as in appendix~\ref{app:numerics}.
They gave
\begin{equation}
 \kappa_{\rm pilot}=1.05(0),\qquad
 \kappa_{\rm refined}=1.09(3) .
\end{equation}
The refined coefficients were approximately
\begin{align}
 \eta&=0.459(2),\qquad
 (a_0,a_1,a_2)=(-0.282(2),-0.0915(0),1.40(4)),\nonumber\\
 (b_0,b_1,b_2)&=(-17.8(8),16.2(3),-4.42(2)).
\end{align}
These rounded values document discovery; the numerical solutions used the
unrounded checkpoint coefficients.
At fixed $\eta$, multiplying all $a_n,b_n$ by a common continuation
parameter from one to zero led to the quadratic model with
$\kappa=1.01(3)$, before continuation to $\eta=1/2$.

\FloatBarrier
\bibliographystyle{JHEP}
\bibliography{references}
\end{document}